# Addestramento con Dataset Sbilanciati

## Addestramento di modelli con l'utilizzo di dataset con classi sbilanciate

Massimiliano Morrelli[1]

[1]Politecnico di Bari - Via Edoardo Orabona, 4, 70126 Bari BA

**Abstract**

**English.** The following document pursues the objective of comparing some useful methods to balance a dataset and obtain a trained model. The dataset used for training is made up of short and medium length sentences, such as simple phrases or extracts from conversations that took place on web channels. The training of the models will take place with the help of the structures made available by the Apache Spark framework, the models may subsequently be useful for a possible implementation of a solution capable of classifying sentences using the distributed environment, as described in "New frontier of textual classification: Big data and distributed calculation" by Massimiliano Morrelli et al.

**Italiano.** Il seguente documento persegue l'obiettivo di mettere a confronto alcuni metodi utili a bilanciare un dataset e ottenere un modello addestrato. Il dataset utilizzato per l'addestramento è composto da frasi di lunghezza breve e media, come frasi semplici o estratte da conversazioni avvenute su canali web.// L'addestramento dei modelli avverrà con l'ausilio delle strutture messe a disposizione dal framework Apache Spark, i modelli successivamente potranno essere utili a un eventuale implementazione di una soluzione in grado di classificare frasi sfruttando l'ambiente distribuito, come descritto in "*Nuova frontiera della classificazione testuale: Big data e calcolo distribuito*" di Massimiliano Morrelli et al.

# Indice

# Elenco delle figure

# 1 Introduzione

Lo studio dei dati con classi sbilanciate è un argomento di notevole importanza, soprattutto nella medicina, nella finanze, nella sicurezza stradale ed altri campi. In presenza di una distribuzione della variabile di risposta estremamente sbilanciata il processo di apprendimento può essere distorto, perché il modello tende a focalizzarsi sulla classe prevalente e ignorare gli eventi rari, che possono essere pazienti aventi un cancro, incidenti stradali mortali, oppure cattivi creditori. Lo studio effettuato in questo documento si basa sull'addestramento di modelli utili alla classificazione delle classe di appartenenza di frasi di lunghezza breve e media come, frasi semplici o estratte da conversazioni avvenute su canali web.// Nei capitoli seguenti, si affrontano le difficoltà che si possono incontrare nel addestrare un modello partendo da un dataset con classi sbilanciate.

# 2 Stato dell'arte

Diverse sono le soluzioni che sono state proposte nel tempo per affrontare il problema dei dati estremamente squilibrati, e si possono distinguere due approcci comuni, tecniche di Cost-Sensitive Learning e tecniche di campionamento. A differenza dei modelli tradizionali di apprendimento, le tecniche Cost-Sensitive utilizzano una funzione di costo di errata classificazione per pesare le diverse classi di risposta e così limitare gli effetti dovuti allo sbilanciamento della distribuzione delle classi stesse. L'obiettivo dell'apprendimento Cost-Sensitive è minimizzare i costi di errata classificazione pesati sulla base di una funzione di penalità.
Le tecniche di campionamento effettuano un lavoro di pre-processing sui dati, in modo da fornire una distribuzione bilanciata tra le classi.
L'uso di metodi di campionamento consiste nella modifica di un set di dati sbilanciati attraverso alcuni meccanismi in modo da fornire una distribuzione equilibrata. Le tecniche più comuni sono l'oversampling che attua un campionamento con ripetizione delle osservazioni appartenenti alla classe rara e l'undersampling che, al contrario, effettua un campionamento senza ripetizione tra le osservazioni appartenenti alla classe maggioritaria.
In altre parole, il random oversampling è un metodo che mira a bilanciare la distribuzione di classe attraverso la replicazione casuale di esempi appartenenti alla classe minoritaria.
Diversi autori concordano sul fatto che l'oversampling può aumentare la probabilità che si verifichino problemi di overfitting.[4]

# 3 Dataset Sbilanciati

Un dataset non bilanciato si verifica quando esiste una differenza molto elevata tra i valori positivi e negativi.
La classificazione sbilanciata si riferisce a un problema di modellazione predittiva della classificazione in cui il numero di esempi nel set di dati di addestramento per ciascuna etichetta di classe non è bilanciato e quindi la sua distribuzione non è uniforme.

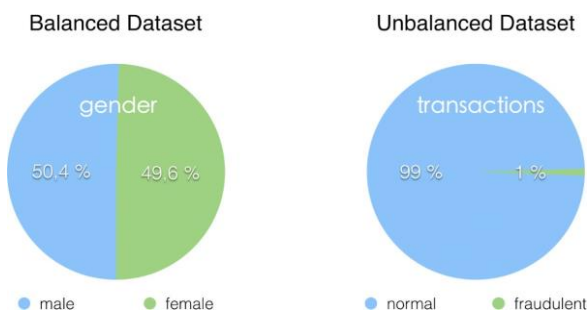

Figura 1: Esempio di dataset bilanciato e sbilanciato

È possibile che lo squilibrio negli esempi tra le classi sia stato causato dal modo in cui gli esempi sono stati raccolti o campionati dal dominio problematico. Ciò potrebbe comportare distorsioni introdotte durante la raccolta dei dati ed errori commessi durante la raccolta dei dati:

- Campionamento parziale.
- Errori di misurazione.

## 3.1 Tecniche di ricampionamento

Le soluzioni basate sui metodi di ricampionamento consistono in un pre-trattamento dei dati, che, come tale, ha il vantaggio di essere indipendente da qualsiasi modello di classificazione e adattabile così a molti contesti diversi. L'obiettivo è quello di alterare preliminarmente la distribuzione delle classi così da alleviarne il grado di sbilanciamento.

### 3.1.1 OverSampling

Questa tecnica viene utilizzata per modificare le classi di dati disuguali, creando un dataset bilanciato. Quando la quantità di dati è insufficiente, il metodo di sovra-campionamento cerca di bilanciare aumentando la dimensione dei campioni rari.
Un campionamento eccessivo aumenta il numero di membri della classe di minoranza nel set di addestramento.
Il vantaggio del sovra-campionamento è che non si perdono informazioni dal set di partenza, poiché vengono mantenute tutte le osservazioni delle classi minoritarie e maggioritarie.

### 3.1.2 UnderSampling

A differenza del sovra-campionamento, questa tecnica bilancia il dataset non bilanciato riducendo la dimensione della classe che è in abbondanza. Il sotto-campionamento, contrariamente al sovra-campionamento, mira a ridurre il numero di campioni di maggioranza per bilanciare la distribuzione delle classi. Poiché rimuove le osservazioni dal set di dati originale, potrebbe scartare informazioni utili.

### 3.1.3 Cost-Sensitive

Cost-Sensitive Learning (CSL) prende in considerazione i costi di classificazione errata minimizzando il costo totale. L'obiettivo di questa tecnica è principalmente quello di perseguire un'elevata precisione di classificazione degli esempi in un insieme di classi conosciute. Questa tecnica viene molto utilizzata negli algoritmi di machine learning tra cui le applicazioni di data mining del mondo reale.

## 3.2 Distribuzione delle classi nel dataset

Di seguito l'andamento della distribuzione delle classi nel dataset utilizzato nel nostro studio.

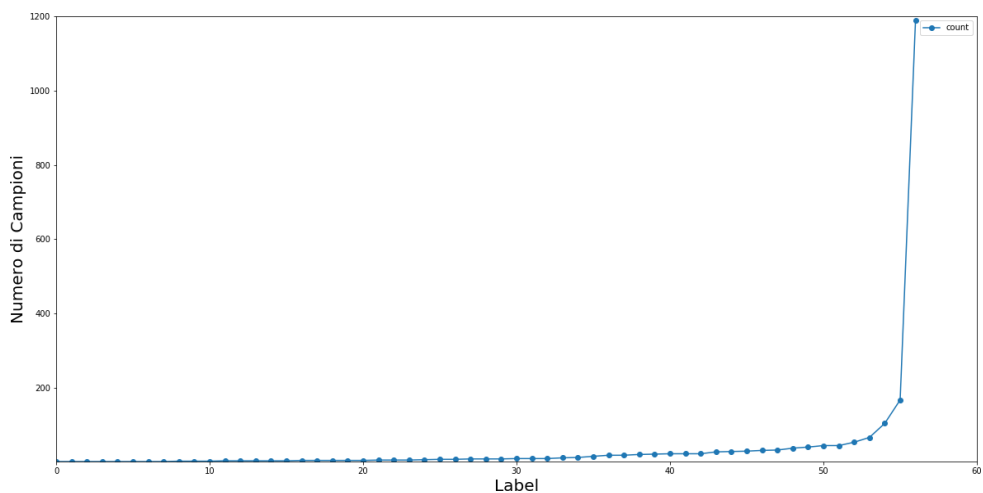

Figura 2: Distribuzione del dataset utilizzato per l'addestramento

Come si evince dalla Figura 2, si nota un andamento crescente e soprattutto una differenza fra la classe maggioritaria che ha una popolosità di 1190 rispetto alle classi minoritarie che hanno una popolosità che non superano i 200 campioni.

# 4 Misure di performance

Un problema legato alle classi di risposta sbilanciate riguarda la valutazione dell'accuratezza del classificatore, ed emerge sia nella scelta della misura dell'errore sia nella stima di esso. Le consuete misure di accuratezza, come ad esempio il tasso di errata classificazione, possono condurre a risultati fuorvianti perché dipendono fortemente dalla distribuzione di classe.

Di seguito vengono sviluppati gli andamenti delle accuracy suddivise per metodo di classificazione, e di conseguenza si può notare dalle relative tabelle che si ottiene un accuracy molto alta.

## 4.1 Distribuzione delle accuracy: Naive Bayes

| *filter* | *accuracy* |
|---|---|
| 10 | 0.74 |
| 15 | 0.60 |
| 20 | 0.65 |
| 25 | 0.70 |
| 30 | 0.83 |
| 35 | 0.77 |
| 40 | 0.87 |
| 45 | 0.75 |
| 50 | 0.88 |

Tabella 1: Accuracy NB

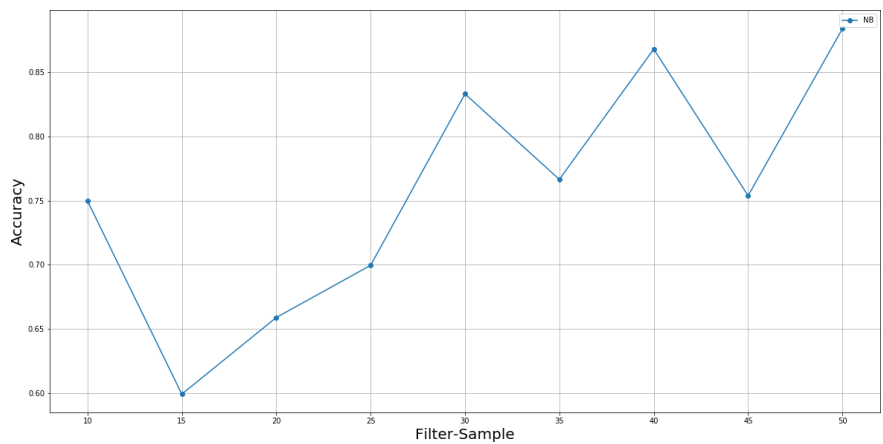

Figura 3: Grafico del andamento delle accuracy: NB

## 4.2 Distribuzione delle accuracy: Support Vector Machine

| filter | accuracy |
|:---:|:---:|
| 10 | 0.80 |
| 15 | 0.64 |
| 20 | 0.70 |
| 25 | 0.73 |
| 30 | 0.88 |
| 35 | 0.82 |
| 40 | 0.93 |
| 45 | 0.79 |
| 50 | 0.94 |

Tabella 2: Accuracy SVM

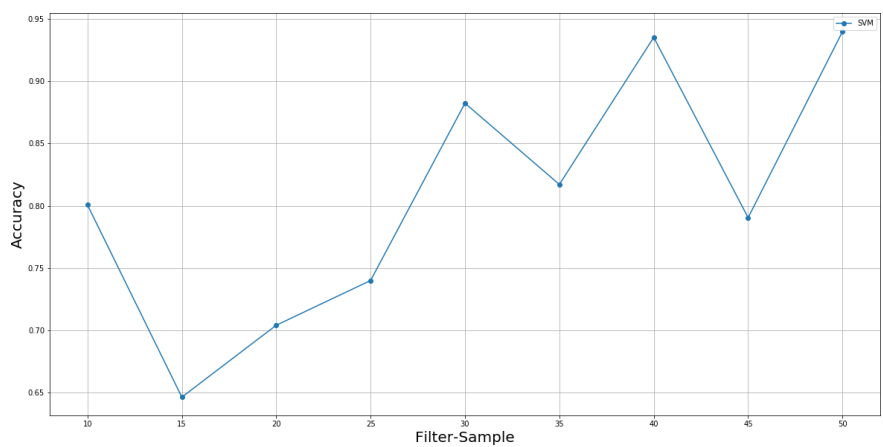

Figura 4: Grafico del andamento delle accuracy: SVN

## 4.3 Distribuzione delle accuracy: Random Forest

| filter | accuracy |
|:---:|:---:|
| 10 | 0.77 |
| 15 | 0.59 |
| 20 | 0.64 |
| 25 | 0.71 |
| 30 | 0.85 |
| 35 | 0.80 |
| 40 | 0.93 |
| 45 | 0.78 |
| 50 | 0.92 |

Tabella 3: Accuracy RF

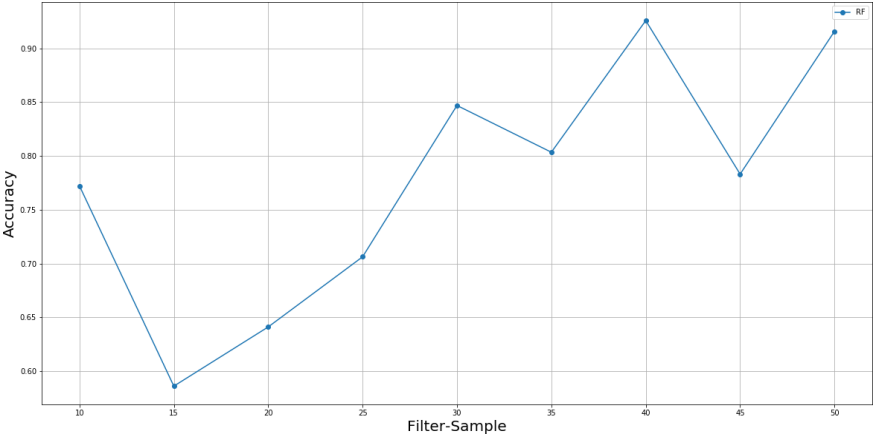

Figura 5: Grafico del andamento delle accuracy: RF

## 4.4 Distribuzione delle accuracy: Deep Learning

| filter | accuracy |
|:---:|:---:|
| 10 | 0.66 |
| 15 | 0.65 |
| 20 | 0.77 |
| 25 | 0.71 |
| 30 | 0.74 |
| 35 | 0.75 |
| 40 | 0.81 |
| 45 | 0.77 |
| 50 | 0.75 |

Tabella 4: Accuracy DNN

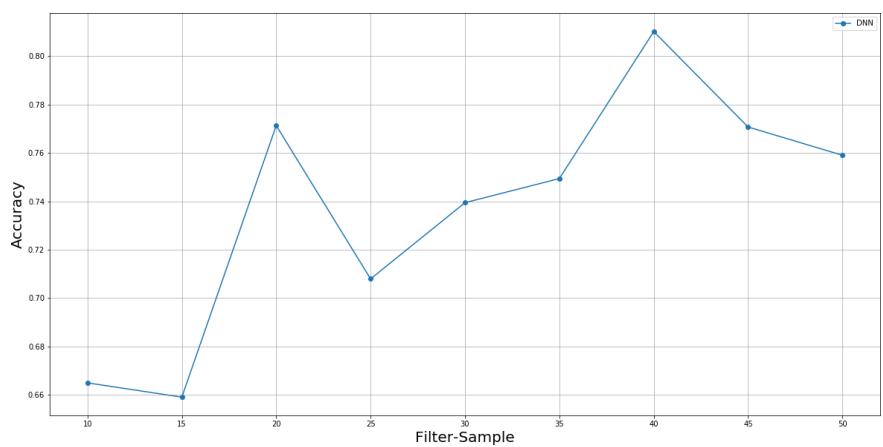

Figura 6: Grafico del andamento delle accuracy: DNN

## 4.5 Matrice di confusione

Il modo più semplice per valutare le prestazioni dei classificatori si basa sull'analisi della matrice di confusione. La matrice di confusione è la matrice contenente le informazioni circa lo stato della realtà e la classificazione ottenuta.

Una formalizzazione generale di una matrice di questo tipo è proposta nella Tabella 5, le righe della matrice sono classi reali, e le colonne sono le classi previste. L'elemento sulla riga i e sulla colonna j è il numero di casi in cui il classificatore ha classificato la classe "vera" j come classe i. Attraverso questa matrice è osservabile se vi è "confusione" nella classificazione di diverse classi.[4]

|     | 1        | 2        | ... | j        | ... | k        |
|-----|----------|----------|-----|----------|-----|----------|
| 1   | $n_{11}$ | $n_{12}$ | ... | $n_{1j}$ | ... | $n_{1k}$ |
| 2   | $n_{21}$ | $n_{22}$ | ... | $n_{2j}$ | ... | $n_{2k}$ |
| ... | ...      | ...      | ... | ...      | ... | ...      |
| j   | $n_{j1}$ | $n_{j2}$ | ... | $n_{jj}$ | ... | $n_{jk}$ |
| ... | ...      | ...      | ... | ...      | ... | ...      |
| k   | $n_{k1}$ | $n_{k2}$ | ... | $n_{kj}$ | ... | $n_{kk}$ |

Tabella 5: Matrice di confusione $k \times k$

Si evince che tutte le previsioni corrette si trovano nella diagonale principale della matrice, quindi gli errori di previsione possono essere facilmente trovati, poiché saranno rappresentati da valori esterni alla diagonale.

### 4.5.1 Metriche base utilizzate dalla matrice di confusione

Per comprendere meglio l'incrocio tra valori effettivi e previsti, alla matrice di confusione vengono associati i seguenti termini:

- **True positive** (TP, o veri positivi): sono i valori presenti nella diagonale principale

- **True negative** (TN, o veri negativi): per ogni classe è la somma di tutti i valori nella matrice di confusione escludendo la riga e la colonna di quella classe.

- **False positive** (FP, o falsi positivi): è la somma di tutti i valori nella colonna escluso il valore di TP.

- **False negative** (FN, o falsi negativi): è la somma di tutti i valori nella riga corrispondente escluso il valore TP.

Tra le metriche base e più diffuse della matrice di confusione possiamo trovare:

**Error Rate**: È il tasso di errore (ERR) e viene calcolato come il numero di tutti i pronostici errati diviso per il numero totale del set di dati. Il miglior tasso di errore è 0, mentre il peggiore è 1.

$$ERR = \frac{FP + FN}{TN + FP + FN + TP} \qquad (1)$$

**Accuracy**: L'accuratezza del modello come dice il nome. Pertanto, la migliore accuratezza è 1, mentre la peggiore è 0. Può anche essere calcolato da 1 – ERR, o dalla seguente formula:

$$Accuracy = \frac{TP + TN}{TN + FP + FN + TP} \qquad (2)$$

**Precision**: La precisione è l'abilità di un classificatore di non etichettare un'istanza positiva che è in realtà negativa. Per ogni classe è definito come il rapporto tra veri positivi e la somma di veri e falsi positivi. Detto in un altro modo, "per tutte le istanze classificate come positive, quale percentuale era corretta?". La formula di calcolo della precisione è la seguente:

$$Precision = \frac{TP}{TP + FP} \qquad (3)$$

**Recall**: La metrica Recall detta anche sensitivity o true positive, è la capacità di un classificatore di trovare tutte le istanze positive. Per ogni classe è definito come il rapporto tra i veri positivi e la somma dei veri positivi e dei falsi negativi. Detto in altri termini, "per tutte le istanze che erano effettivamente positive, quale percentuale è stata classificata correttamente?". La formula di calcolo del richiamo è la seguente:

$$Recall = \frac{TP}{TP + FN} \qquad (4)$$

### 4.5.2 Analisi della Matrice di Confusione senza campionamento

Per il calcolo della matrice di confusione abbiamo selezionato il valore di filter che restituiva un accuracy media più alta. Di seguito le accuracy per metodo.

|    | 1   | 2    | 3     | 4   | 5   | 6   | 7   | 8    | 9   | 10  | 11   | 12  | 13   | 14   |
|----|-----|------|-------|-----|-----|-----|-----|------|-----|-----|------|-----|------|------|
| 1  | 3.0 | 0.0  | 1.0   | 0.0 | 0.0 | 0.0 | 0.0 | 0.0  | 0.0 | 0.0 | 0.0  | 0.0 | 0.0  | 0.0  |
| 2  | 0.0 | 12.0 | 8.0   | 0.0 | 0.0 | 0.0 | 0.0 | 2.0  | 0.0 | 0.0 | 1.0  | 0.0 | 0.0  | 0.0  |
| 3  | 0.0 | 1.0  | 428.0 | 1.0 | 3.0 | 0.0 | 1.0 | 1.0  | 0.0 | 0.0 | 4.0  | 0.0 | 1.0  | 5.0  |
| 4  | 0.0 | 0.0  | 5.0   | 5.0 | 0.0 | 0.0 | 0.0 | 0.0  | 0.0 | 0.0 | 0.0  | 0.0 | 1.0  | 0.0  |
| 5  | 0.0 | 1.0  | 5.0   | 0.0 | 9.0 | 0.0 | 0.0 | 0.0  | 0.0 | 0.0 | 1.0  | 0.0 | 0.0  | 0.0  |
| 6  | 0.0 | 0.0  | 7.0   | 0.0 | 0.0 | 3.0 | 0.0 | 0.0  | 0.0 | 0.0 | 0.0  | 0.0 | 0.0  | 0.0  |
| 7  | 0.0 | 0.0  | 5.0   | 0.0 | 0.0 | 0.0 | 8.0 | 0.0  | 0.0 | 0.0 | 2.0  | 0.0 | 0.0  | 0.0  |
| 8  | 0.0 | 0.0  | 14.0  | 0.0 | 0.0 | 0.0 | 0.0 | 15.0 | 0.0 | 0.0 | 0.0  | 0.0 | 0.0  | 0.0  |
| 9  | 0.0 | 0.0  | 7.0   | 0.0 | 0.0 | 0.0 | 0.0 | 0.0  | 5.0 | 0.0 | 0.0  | 0.0 | 0.0  | 0.0  |
| 10 | 0.0 | 0.0  | 5.0   | 0.0 | 0.0 | 0.0 | 0.0 | 0.0  | 0.0 | 7.0 | 1.0  | 0.0 | 0.0  | 1.0  |
| 11 | 0.0 | 0.0  | 34.0  | 0.0 | 0.0 | 0.0 | 0.0 | 0.0  | 0.0 | 0.0 | 24.0 | 0.0 | 0.0  | 0.0  |
| 12 | 0.0 | 0.0  | 9.0   | 0.0 | 0.0 | 0.0 | 0.0 | 0.0  | 0.0 | 0.0 | 0.0  | 2.0 | 0.0  | 0.0  |
| 13 | 0.0 | 0.0  | 9.0   | 0.0 | 0.0 | 0.0 | 0.0 | 0.0  | 0.0 | 0.0 | 0.0  | 0.0 | 11.0 | 0.0  |
| 14 | 0.0 | 0.0  | 22.0  | 0.0 | 0.0 | 0.0 | 0.0 | 0.0  | 0.0 | 0.0 | 2.0  | 0.0 | 0.0  | 23.0 |

Tabella 6: Matrice di confusione: NB

Come si evince dalla matrice di confusione nella Tabella 6, notiamo che la terza colonna, in blu, è piena e quindi si intuisce che le frasi vengono classificate come appartenenti alla classe 3 anche se appartengono ad altre classi. Inoltre si nota che il TP della classe 3 è molto alto e nella sua riga ci sono pochi FN quindi si può dedurre che il valore altro di accuracy ottenuta dal modello è falsata dalla classe maggioritaria.

# 5 Implementazione del MiddleSample

Per rendere uniforme la distribuzione del dataset si è deciso di implementare un metodo ibrido che permette di sfruttare i vantaggi di entrambi i metodi di ricampionamento trattati nel Cap. 3.

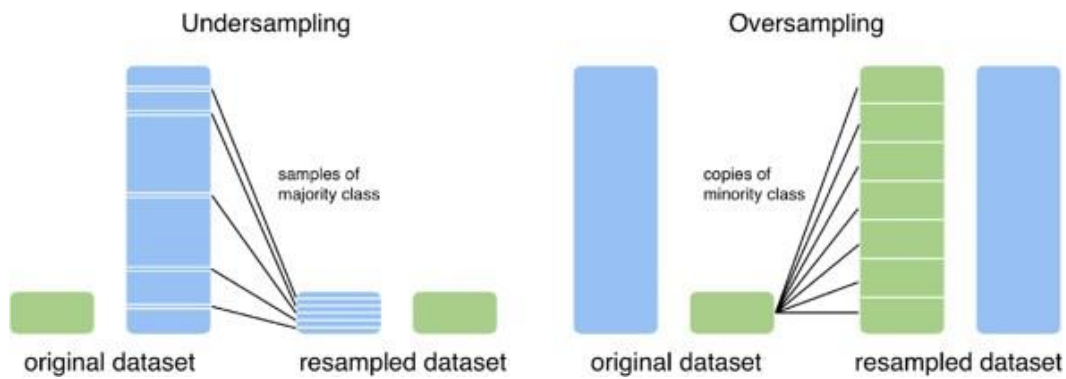

Figura 7: Sottocampionamento e Sovracampionamento

Il metodo **MiddleSample** effettua prima un operazione di filtraggio e successivamente esegue il bilanciamento dei campioni.
La prima operazione, ovvero quella di filtraggio, avviene tramite un parametro, *filter*, che seleziona solo quelle classi che hanno una popolosità pari o maggiore al valore di filter. Questa operazione viene effettuata per escludere ad esempio classi che hanno una popolosità così bassa i quali non basterebbero per addestrare un modello, anche se il dataset fosse bilanciato.
Successivamente si effettua il ricampionamento, ovvero tramite un valore di *sample*, si sovracampioneranno le classi minoritarie verranno sotto-campionate le classi maggioritarie.

## 5.1 Implementazione in Apache Spark

Per poter sfruttare la potenza del calcolo distribuito è stato utilizzato il framework di Apache Spark nella versione 2.3.1.
L'implementazione del MiddleSample si basa su un ciclo che sovra-campiona tutte le classi e successivamente estrapola in maniera stocastica il numero di campioni definiti in sample.

```java
public Dataset<Row> getSamplingData(Dataset<Row> df, int sample){
 List<Dataset<Row>> datasetList = new ArrayList<Dataset<Row>>();
 Utility.getCategoryList(df).forEach(label -> {
  Dataset<Row> _df = df.filter(functions.col("label").equalTo(label));
  Long replicationValue = 0L;
  replicationValue = sample/_df.count() + 1;
  Integer[] nRepl = new Integer[replicationValue.intValue()];
  Arrays.fill(nRepl, 0);
  Dataset<Row> dfRepl = _df.agg(functions.collect_list("occurrence"))
    .withColumnRenamed("collect_list(occurrence)", "occurrencies")
    .withColumn("nRepl", functions.lit(nRepl))
    .withColumn("repl", functions.explode(functions.col("nRepl")))
    .withColumn("occurrence", functions.explode(functions.col("occurrencies")))
    .withColumn("label", functions.lit(label))
    .select(" label"," occurrence")
    .orderBy( functions.rand())
    .limit(sample);
  datasetList.add(dfRepl);
 });
 Iterator<Dataset<Row>> iterDataset = datasetList.iterator();
 Dataset<Row> _df = iterDataset.next();
 while (iterDataset.hasNext()) {
  _df = _df.union(iterDataset.next());
 }
 return _df;
}
```

# 6 Risultati ottenuti

In questo capitolo tratteremo gli andamenti delle accuracy in base ai valori imposti di sample e filter.

## 6.1 Andamento delle accuracy

Di seguito è possibile analizzare gli andamenti delle accuracy ottenute con l'addestramento dei modelli descritti in *"Nuova frontiera della classificazione testuale:Big data e calcolo distribuito"*
I valori delle accuracy sono stati ottenuti variando i parametri di $filter \in [0, 50]$ con un incremento di 5 e $sample \in [100, 1000]$ con un passo pari a 50.

### 6.1.1 Distribuzione delle accuracy: Naive Bayes

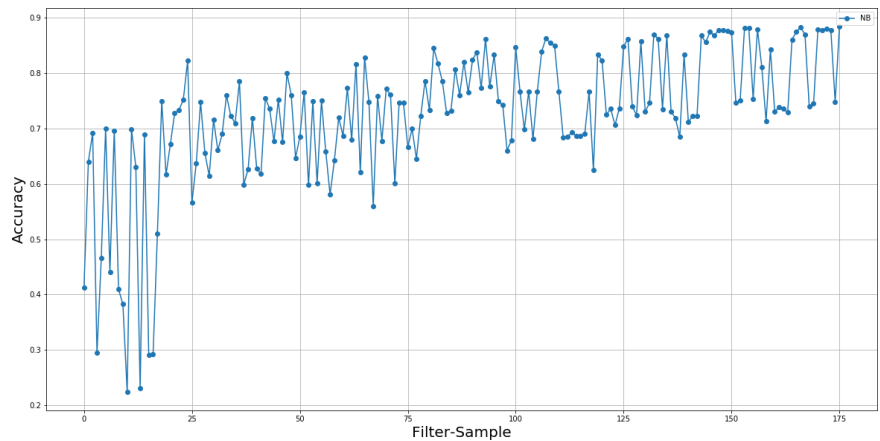

Figura 8: Grafico del andamento delle accuracy: NB

### 6.1.2 Distribuzione delle accuracy: Support Vector Machine

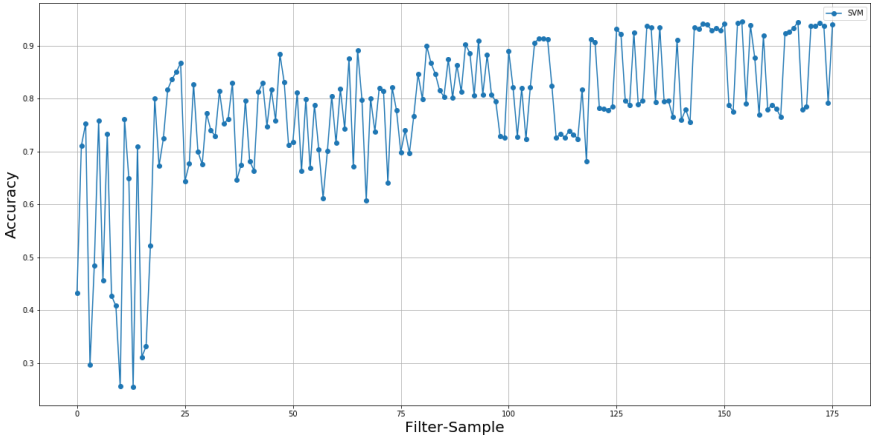

Figura 9: Grafico del andamento delle accuracy: SVM

### 6.1.3 Distribuzione delle accuracy: Random Forest

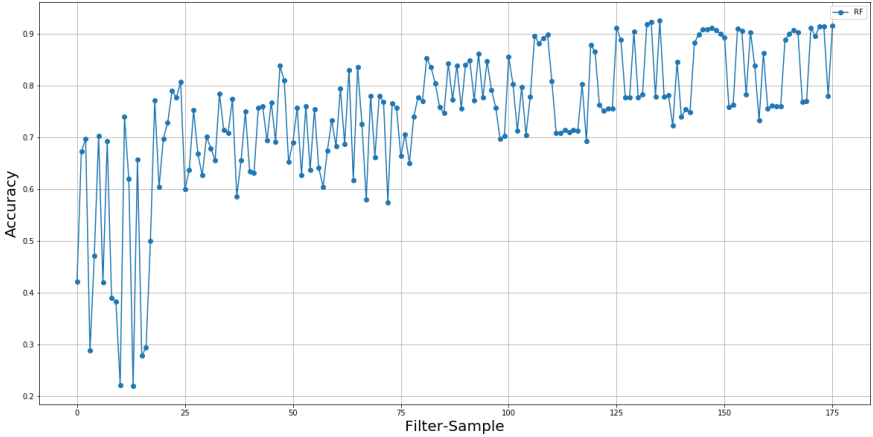

Figura 10: Grafico del andamento delle accuracy: RF

### 6.1.4 Distribuzione delle accuracy: Deep Learning

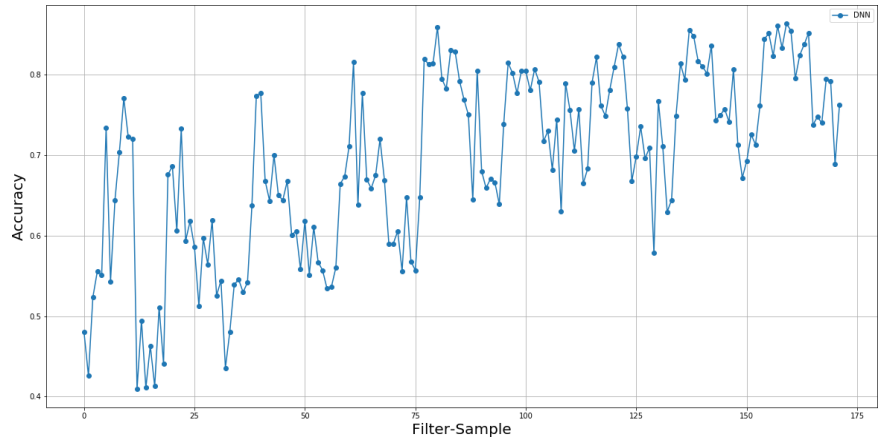

Figura 11: Grafico del andamento delle accuracy: DNN

### 6.1.5 Analisi della Matrice di Confusione con il metodo MiddleSample

Come si può evincere dalla matrice descritta nella Tabella 7 notiamo che la diagonale principale risulta più popolata.

|   | 1 | 2 | 3 | 4 | 5 | 6 | 7 | 8 | 9 | ∑ | Recall |
|---|---|---|---|---|---|---|---|---|---|---|---|
| 1 | **11.0** | **0.0** | 0.0 | 0.0 | 0.0 | 0.0 | 1.0 | 0.0 | 0.0 | 12.0 | 0.91 |
| 2 | 0.0 | **262.0** | 0.0 | 0.0 | 2.0 | 0.0 | 0.0 | 1.0 | 1.0 | 266.0 | 0.98 |
| 3 | 0.0 | **1.0** | **6.0** | 0.0 | 0.0 | 0.0 | 0.0 | 0.0 | 0.0 | 7.0 | 0.86 |
| 4 | 0.0 | **1.0** | 0.0 | **4.0** | 0.0 | 0.0 | 0.0 | 0.0 | 0.0 | 5.0 | 0.8 |
| 5 | 0.0 | **0.0** | 0.0 | 0.0 | **9.0** | 0.0 | 0.0 | 0.0 | 0.0 | 9.0 | 1.0 |
| 6 | 0.0 | **2.0** | 0.0 | 0.0 | 0.0 | **9.0** | 0.0 | 0.0 | 0.0 | 11.0 | 0.82 |
| 7 | 0.0 | **11.0** | 0.0 | 0.0 | 0.0 | 0.0 | **24.0** | 0.0 | 0.0 | 35.0 | 0.69 |
| 8 | 0.0 | **0.0** | 0.0 | 0.0 | 0.0 | 0.0 | 0.0 | **7.0** | 0.0 | 7.0 | 1.0 |
| 9 | 0.0 | **1.0** | 0.0 | 0.0 | 0.0 | 0.0 | 0.0 | 0.0 | **23.0** | 24.0 | 0.96 |

Tabella 7: Matrice di confusione con campionamento ibrido

Per poter paragonare i miglioramenti ottenuti utilizzando il metodo di MiddleSample è necessario calcolare i valori di Recall della Tabella 6

|    |       | Recall |
|----|-------|--------|
| 1  | 4.0   | 0.75   |
| 2  | 23.0  | 0.52   |
| 3  | 445.0 | 0.96   |
| 4  | 11.0  | 0.45   |
| 5  | 16.0  | 0.56   |
| 6  | 10.0  | 0.3    |
| 7  | 15.0  | 0.53   |
| 8  | 39.0  | 0.38   |
| 9  | 12.0  | 0.58   |
| 10 | 14.0  | 0.5    |
| 11 | 58.0  | 0.41   |
| 12 | 11.0  | 0.2    |
| 13 | 20.0  | 0.55   |
| 14 | 47.0  | 0.49   |

Tabella 8: Calcolo del valore della Recall

Dopo aver calcolato il valore di Recall di entrambe le matrici di confusioni si calcola il valore medio delle recall e cosi facendo notiamo che la media della recall della matrice senza campionamento è $\mu_{recall} = 0.51$ mentre la media ottenuta utilizzando i valori della Tabella 7 è $\mu_{recallMS} = 0.89$.

### 6.2 Cost-Sensitive: implementazione nel deeplearning

In letteratura si evince che il metodo Cost-Sensitive è più adatto per le reti neurali e di conseguenza abbiamo provato ad addestrare il modello con il suddetto metodo, per l'analisi delle accuracy.

```
MultiLayerConfiguration conf = new NeuralNetConfiguration.Builder()
  .seed(Const.DeepProperty.SEED)
  .optimizationAlgo(OptimizationAlgorithm.STOCHASTIC_GRADIENT_DESCENT)
  .updater(new Adam())
  .list()
  .layer(0, new DenseLayer.Builder().activation(Activation.ELU)
    .nIn(numInputs)
    .nOut(Const.DeepProperty.UNIT)
    .build())
  .layer(1, new OutputLayer.Builder(new LossMCXENT(weightsArray))
    .activation(Activation.SOFTMAX)
    .nIn(Const.DeepProperty.UNIT)
    .nOut(outputNum)
    .build())
  .build();
```

L'array *weightsArray* contiene i pesi delle categorie, ordinati per posizione e calcolati secondo la seguente formula:

$$w_{label} = 1 - \frac{\Sigma x_{label}}{\Sigma x_{dataset}} \qquad (5)$$

### 6.2.1 Distribuzione delle accuracy: Deep Learning

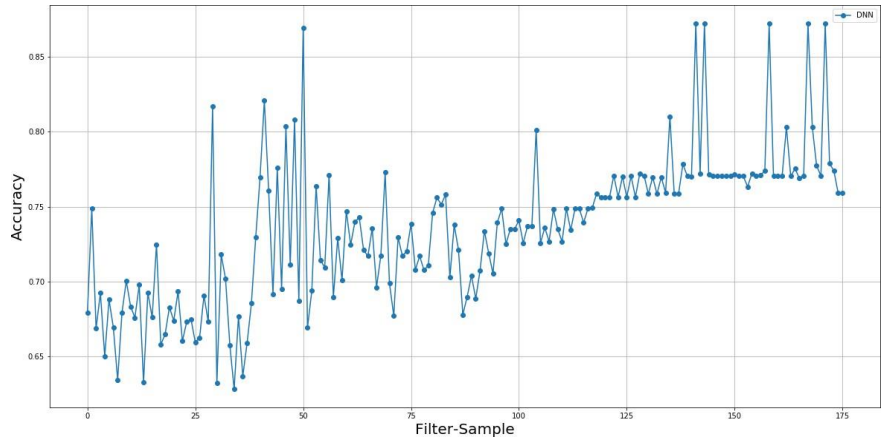

Figura 12: Grafico del andamento delle accuracy: DNN

Come si può notare dalla figura 12 l'andamento delle accuracy risulta inferiore rispetto a quello descritto nella figura 11, anche in questo caso risulta più accurato l'uso dell'middle sample.

# 7 Conclusioni

Dall'analisi della letteratura è stato osservato che in presenza di una distribuzione della variabile di risposta estremamente sbilanciata il processo di apprendimento può essere distorto. Le tecniche più comuni sono l'oversampling e l'undersampling.

Nel presente lavoro è stata presentata una nuova metodologia, il ricampionamento *MiddleSample* e analizzando le distribuzioni degli andamenti delle accuracy nel Capitolo 6, si può dedurre che utilizzando il middleSample si migliora il grado di accuratezza dell'addestramento mantenendo una matrice di confusione con valori TP abbastanza alti.

L'utilizzo del metodo di ricampionamento MiddleSample ci permette di avere un accuratezza massima per metodo pari a:

- NAIVE BAYES: 0.85
- SUPPORT VECTOR MACHINE: 0.91
- RANDOM FOREST: 0.89
- DEEP LEARNING: 0.75

con i parametri, *filter* = 35, e *sample* = 600

## 7.1 Area di miglioramento e possibili evoluzioni

Si evince anche, che, più i coefficienti *filter* e *sample* si avvicina all'estremo destro, più si effettuerà un azione di taglio dei campioni utili all'addestramento, e quindi si vanno ad escludere la maggior parte delle classi, e di conseguenza informazioni. Quindi si potrebbe valutare di integrare un ulteriore parametro, come ad esempio il numero di classi utilizzate per addestrare il modello, che permetta di limitare l'esclusione di informazioni.

Una possibile evoluzione, nel caso della classificazione di testi, sarebbe l'inserimento di un pre-trattamento del testo così da ridurre lo spazio vettoriale per l'addestramento e inoltre andare a semplificare quelle frasi che per la loro molteplicità di significato appartengono a più classi.

# Bibliografia